\documentclass[useAMS,usenatbib]{mn2e_v2}
\usepackage{amsmath,graphics,epsfig}

\title[Compact broadband planar orthomode transducer]{Compact broadband planar orthomode transducer}
\author[P.K. Grimes, O.G. King, G. Yassin and M.E. Jones]{P.K. Grimes\thanks{}, O.G. King, G. Yassin and M.E. Jones\\
Dept. of Physics, University of Oxford, Denys Wilkinson Building, Keble Road, Oxford OX1 3RH, UK\\}

\begin{document}
\date{\today}
\pagerange{\pageref{firstpage}--\pageref{lastpage}} \pubyear{2007}
\maketitle

\begin{abstract}
We present the design and test results of a compact C-band orthomode transducer which comprises four rectangular probes orthogonally arranged in a circular waveguide, designed to work in the WG13 band. Measurements of the system in the frequency range 4.64~GHz to 7.05~GHz agree very well with simulation results and show a cross-polarisation level below -58~dB, a return loss of about -20~dB, and an insertion loss difference of less than 0.18~dB between the orthogonal polarisation modes across the full waveguide band.
\end{abstract}

\footnotetext{Email: pxg@astro.ox.ac.uk\\This paper is a preprint of a paper accepted by Electronics Letters and is subject to Institution of Engineering and Technology Copyright. When the final version is published, the copy of record will be available at IET Digital Library}

\section{Introduction}
An orthomode transducer (OMT) is used to extract two orthogonal polarisation modes from a rectangular or circular waveguide. Several radio and mm cosmological experiments are now being designed to measure weakly polarised sources with an unprecedented sensitivity and accuracy \citep{Taylor:2006}. These instruments require an OMT with extremely low cross-polarisation of order -50~dB or less and near-equal insertion loss of both orthogonal modes over a relatively wide band. The OMT must also be compact, easy to fabricate and integrate with planar circuits, and coolable to 4.2~K.

An L-band OMT design, comprising four probes at right angles in a cylindrical waveguide, proposed by D. Bock \citep{Bock:1999b} has many of these required properties. Two orthogonal polarisations are extracted by combining the signals from each pair of opposite probes. It is compact, simple to construct, intrinsically relatively wideband, and easy to scale to any frequency band. Attempts to realise a high performance OMT using this design have so far fallen short of achieving the stringent requirements of new polarisation cosmology instruments, either because the bandwidth was too narrow \citep{Jackson:2001} or the cross-polarisation was too high \citep{Engargiola:2003}.

\begin{figure}
\caption[Top view of the completed OMT. The waveguide diameter is 41.4~mm, the probe dimensions are 11.46~mm by 4.25~mm, and the backshort length is 14.8~mm]{Top view of the completed OMT. The waveguide diameter is 41.4~mm, the probe dimensions are 11.46~mm by 4.25~mm, and the backshort length is 14.8~mm}
 \centering
 \includegraphics[width=8.6cm,bb=0 0 598 593]{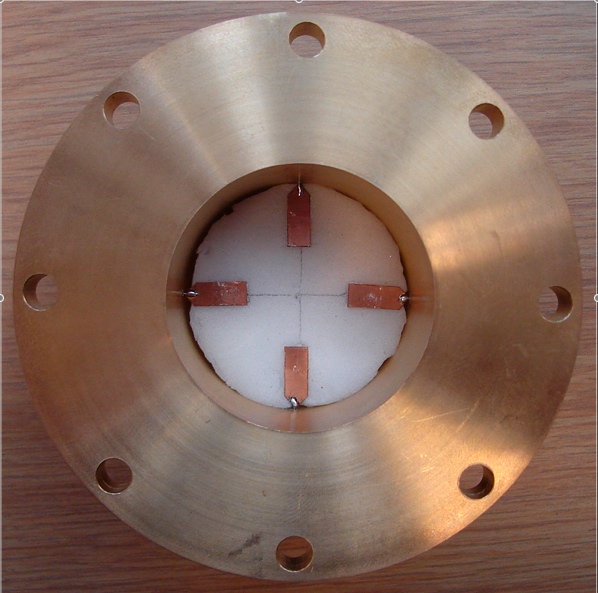}
 \label{fig:omt_pic}
\end{figure}


The OMT described in this paper was developed for the C-Band All-Sky Survey (C-BASS) receiver which is being constructed by the Experimental Radio Cosmology group (ERC) at Oxford. The C-BASS receiver will be used to produce an all-sky map with unprecedented accuracy and sensitivity of the C-band polarised synchrotron radiation from our Galaxy. The data will be used primarily in the subtraction of foregrounds from upcoming measurements of the B-mode polarised component of the Cosmic Microwave Background (CMB) \citep{Taylor:2006}.

\section{Design and Simulation}
The OMT was designed to cover the full WG13 (4.64~GHz to 7.05~GHz) band. The waveguide body was machined from brass and contained four rectangular probes, cut from 0.2~mm thick copper sheeting, placed orthogonally in a single cross-sectional plane of a circular waveguide on a supporting bed of a very low dielectric constant substance ($\simeq$~1.06, Plastazote LD45\footnote{http://www.zotefoams.com/pages/EN/plastazote.asp}), which filled the space between the probes and the backshort (Fig.~\ref{fig:omt_pic}). The probes were tapered at an angle of 45$^{\circ}$ to the waveguide walls to reduce the capacitance and were soldered to the pins of the SMA jacks, which were grounded to the body of the waveguide and fed through the wall in front of a fixed backshort. The signals from each pair of probes were combined by equal-length semi-rigid cables to a 180$^{\circ}$ hybrid, based on a two branch rat-race design \citep{Knochel:1990}, to produce the in-phase orthogonal polarisation signals.

The OMT was modeled with Ansoft's HFSS\footnote{http://www.ansoft.com/products/hf/hfss/} package using thin perfectly conducting probes and 50~$\Omega$ lumped element ports, with the probe dimensions and backshort length optimised for good return loss over a 45\% bandwidth. The optimum OMT waveguide model was combined with the hybrid model in Ansoft Designer\footnote{http://www.ansoft.com/ansoftdesigner/} to produce simulations of the full OMT.

\section{Experimental Results}
\begin{figure}
\caption[Measurement setup showing the Device Under Test - the coax-to-waveguide adaptor, rectangular-to-circular waveguide transition, waveguide straight, and OMT]{Measurement setup showing the Device Under Test - the coax-to-waveguide adaptor, rectangular-to-circular waveguide transition, waveguide straight, and OMT}
 \centering
 \includegraphics[width=8.6cm, bb=0 0 800 1067]{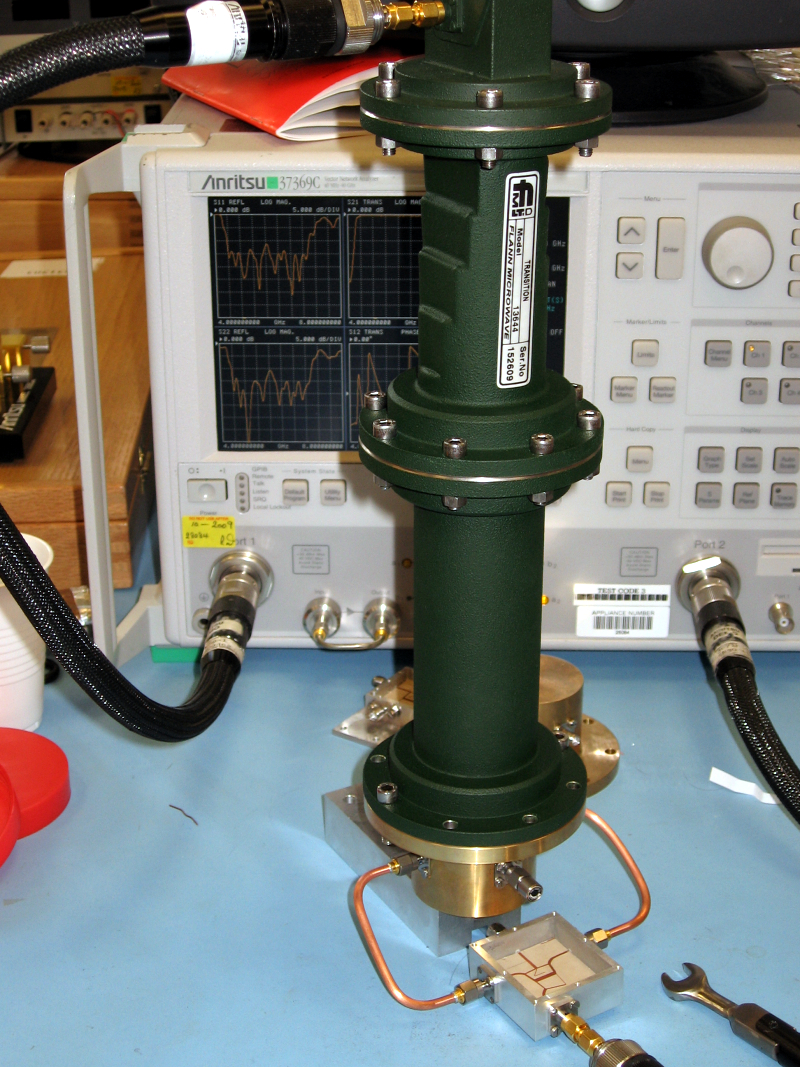}
 \label{fig:OMT_test_setup}
\end{figure}

The system was tested using an Anritsu 37369C Vector Network Analyser (VNA), calibrated to remove the effects of the coaxial cables only. The OMT measurement setup is shown in Fig.~\ref{fig:OMT_test_setup}. Port~1 of the VNA was connected to a coax-to-rectangular waveguide adaptor. This fed a rectangular-to-circular waveguide transition and then a circular waveguide section which was connected to the OMT input. Each polarisation component was measured by terminating the outputs from one pair of opposite probes while the second pair of opposite probes were connected to the hybrid. Port~2 of the VNA was connected to the difference output of the hybrid and the sum output of the hybrid was terminated with a matched load. The circular waveguide section was added to cut off any spurious evanescent cross-polarised modes produced by the rectangular-to-circular transition.

\begin{figure}
 \centering
 \includegraphics[width=8.6cm]{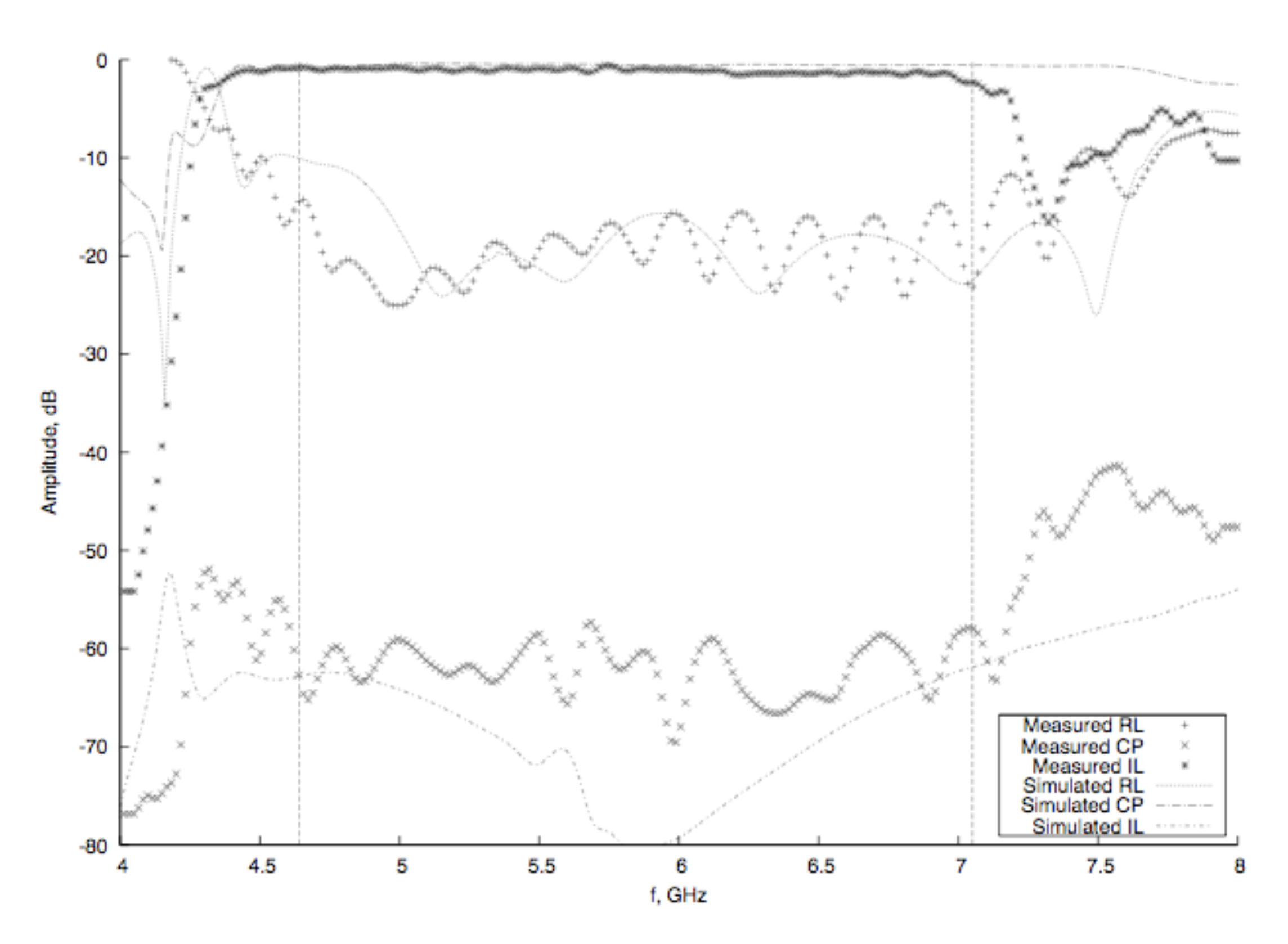}
\caption[Simulated characteristics of the OMT, and the measured response of the Device Under Test in Fig. \protect\ref{fig:OMT_test_setup}. The traces plotted are return loss (RL), insertion loss (IL) and cross-polarisation (CP). The rated band of the waveguide (WG13) is indicated by vertical lines]{Simulated characteristics of the OMT, and the measured response of the Device Under Test in Fig. \protect\ref{fig:OMT_test_setup}. The traces plotted are return loss (RL), insertion loss (IL) and cross-polarisation (CP). The rated band of the waveguide (WG13) is indicated by vertical lines}
 \label{fig:measured_response}
\end{figure}

\begin{figure}
 \centering
 \includegraphics[width=8.6cm]{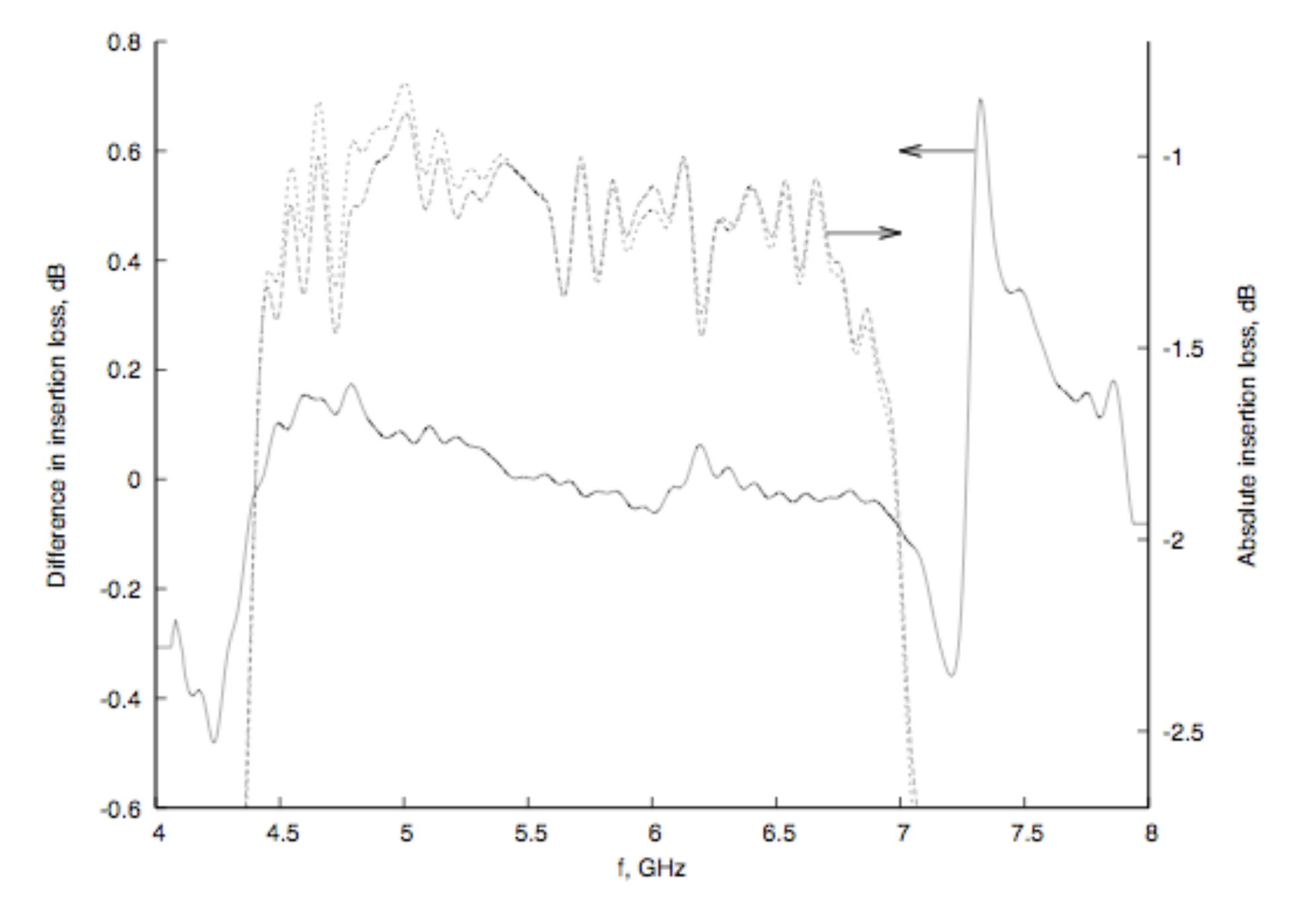}
\caption[Measured insertion loss of both pairs of opposite probes and the difference between them]{Measured insertion loss of both pairs of opposite probes and the difference between them}
 \label{fig:measured_IL_difference}
\end{figure}

The insertion loss and return loss of each pair of probes were measured by aligning the input waveguide with the target probes. The cross-polarisation for each pair of probes was then measured by rotating the input waveguide by 90$^{\circ}$. The measured and simulated response of the OMT are shown in Fig.~\ref{fig:measured_response}. The measured cross-polarisation, return loss, and insertion loss agreed very well with the simulation predictions. The WG13 band of the circular waveguide is indicated by vertical dashed lines. It can be seen that the cross-polarisation and return loss of the OMT, including the coaxial-to-waveguide adaptor and rectangular-to-circular waveguide transition, are about -58~dB and -20~dB respectively. The difference in insertion loss between the two pairs of opposite probes was found to be less than 0.18~dB over the design band, as shown in Fig. \ref{fig:measured_IL_difference}. When incorporated into the C-BASS receiver the OMT will be cooled to 4.2~K, significantly reducing insertion losses. The measurements were repeated for OMT waveguide bodies with 1~mm longer and shorter backshort lengths, with no significant change in performance.

\section{Conclusion}
We have presented a novel design of a compact planar C-band four probe circular waveguide OMT. We have tested the OMT in the frequency range 4.64~GHz to 7.05~GHz and found excellent agreement between simulations and measured results. We have demonstrated a very low cross-polarisation of less than -58~dB, a return loss of about -20~dB, and a difference in insertion loss between the orthogonal modes of less than 0.18~dB over the design band. The OMT is compact, easy to fabricate and yet satisfies the strict requirements of a new generation of radio and mm cosmological polarisation experiments.

\bibliographystyle{apalike_ru}
\pagestyle{plain} 
\bibliography{../ogk.bib}

\begin{thebibliography}{}

\bibitem[Bock, 1999]{Bock:1999b}
Bock, D. (1999), BIMA Memo No. 74.

\bibitem[Engargiola and Plambeck, 2003]{Engargiola:2003}
Engargiola, G. and Plambeck, R. (2003), Review of Scientific Instruments,
  74(3), 1380--1382.

\bibitem[Jackson, 2001]{Jackson:2001}
Jackson, R. (2001), IEEE Microwave and Wireless Components Letters, 11(12),
  483--485.

\bibitem[Kn\"ochel and Mayer, 1990]{Knochel:1990}
Kn\"ochel, R. and Mayer, B. (1990), \IMTTS.

\bibitem[Taylor, 2006]{Taylor:2006}
Taylor, A. (2006), New Astronomy Reviews, 50(11-12), 993--998.

\end{thebibliography}

\end{document}